\documentclass[cup5b]{caps}
\usepackage{psfig}
\usepackage{graphicx}
\usepackage{amssymb}
\usepackage{ociwsymp1e}
\HeadText{E. M. Sadler}

\begin{document}

\pagenumbering{arabic}

\author[]{E. M. SADLER\\School of Physics, University of Sydney, NSW 2006, Australia}

\chapter{The Local Black Hole Mass Function \\ from 2dFGRS Radio Galaxies}

\begin{abstract}

Radio emission from early--type (E and S0) galaxies usually signals the 
presence of a central supermassive black hole.  If radio luminosity and  black--hole mass are correlated, then we can use the large sample of 
radio galaxies observed in the 2dF Galaxy Redshift Survey (2dFGRS) 
to estimate the local ($z<0.1$) black--hole mass function 
for early--type galaxies. 

Integrating over this mass function yields a local black--hole mass density of 
$\rho_{\rm BH}\ =\ 1.8^{+0.4}_{-0.6}\times10^5\ M_\odot\ {\rm Mpc}^{-3}$ 
($H_0$ = 50 km s$^{-1}$ Mpc$^{-1}, \Omega_0$=1), in good agreement 
both with the local value derived from galaxy velocity dispersions and 
with the high--redshift value derived from QSOs. 
This supports earlier suggestions that local radio galaxies 
are the direct descendants of most or all of the high--redshift QSOs. 

\end{abstract}

\section{Does Radio Luminosity Scale with Black  Hole Mass?} 

The relationship between black--hole mass and radio luminosity in 
nearby galaxies is currently a topic of much debate.   
Auriemma et al.\ (1977) first showed that (optically) brighter 
elliptical galaxies are more likely to be radio sources, so the discovery 
of a correlation between black--hole mass and bulge luminosity (Magorrian 
et al.\ 1998) means that some correlation between black--hole mass and 
radio luminosity must exist for early--type galaxies, at least in a 
statistical sense (there may still be considerable scatter for individual objects).  

Franceschini, Vercellone \& Fabian (1998) found a remarkably tight 
relationship between black--hole mass and radio power 
($P_{\rm radio}\propto\,$$M_{\rm BH}^{2.5}$) in nearby galaxies, 
implying that the total radio power emitted by a galaxy is both an 
excellent tracer of the presence of a supermassive black hole and a 
good estimator of its mass.  
However, analysis of larger data sets by Laor (2000) and Ho (2002) found 
a much larger scatter in the $M_{\rm BH}$--$P_{\rm radio}$ correlation.  
All these authors used data sets which contained a mixture of galaxy types 
(e.g. spirals, ellipticals, Seyfert galaxies, radio galaxies and quasars).  

More recently, Snellen et al.\ (2003) have investigated the correlation 
between black--hole mass and radio power for the large sample 
of nearby elliptical galaxies with stellar velocity dispersions measured 
by Faber et al.\ (1989).  They found that that most optically--selected 
(i.e.\ `relatively passive') elliptical galaxies obeyed the Franceschini 
et al.\ (1998) relation between black--hole mass and radio luminosity.  

Black hole mass is clearly not the only variable which determines radio 
luminosity --- we know that radio galaxies were more powerful and/or 
more numerous in the past, whereas the mass of their central black holes 
can only increase with time. Lacy et al.\ (2001) suggest that radio power 
scales with both black--hole mass and accretion rate, which provides a 
plausible mechanism for time--varying radio emission. 
 
\section{The 2dF Galaxy Redshift Survey} 
The recently-completed 2dF Galaxy Redshift Survey (2dFGRS; Colless 
et al.\ 2001) obtained optical spectra and redshifts for over 220,000 
galaxies in the local ($z<$0.3) universe.  Cross-matching these galaxies 
with sensitive all--sky radio imaging surveys like the NRAO VLA Sky Survey 
(NVSS; Condon et al.\ 1998) makes it possible to assemble samples of several 
thousand radio-galaxy spectra which can be used to derive accurate radio 
luminosity functions for both AGN and star-forming galaxies 
(Sadler et al.\ 2002).  

Cross-matching the 2dFGRS and NVSS surveys provides a data set which 
is both large and very homogeneous in quality.  Although only about 
1.5\% of 2dFGRS galaxies are matched with radio sources above the NVSS 
detection limit of 2.5\,mJy at 1.4\,GHz, this still represents one of the 
largest and most uniform set of radio--galaxy spectra so far obtained. 
The high quality of most of the 2dFGRS spectra means that it is usually 
straightforward to determine whether the radio emission arises mainly 
from an AGN, or from processes related to star formation. The final data 
set (work in progress) will be a factor of four larger. 

Most 2dFGRS--NVSS radio galaxies (at least 90\% 
of those classified as AGN) show either an absorption--line spectrum 
typical of giant ellipticals or LINER--like emission lines superimposed 
on a stellar continuum, i.e.\ they fall into the class of `relatively passive' 
radio galaxies discussed by Snellen et al.\ (2003).  It therefore 
seems reasonable to use the relation between black--hole mass and radio 
power from Franceschini et al.\ (1998) together with the accurate local 
radio luminosity function from the 2dFGRS--NVSS sample to estimate the 
local mass function for massive black holes.  This calculation only 
requires that the Franceschini et al.\ (1998) relation holds in a 
statistical sense (there can still be some scatter for individual objects) 
for early--type galaxies with relatively quiescent optical spectra.  

\section{The Local Radio Luminosity Function for 2dFGRS Radio Galaxies} 

\begin{figure}
\centering
\includegraphics[width=\columnwidth]{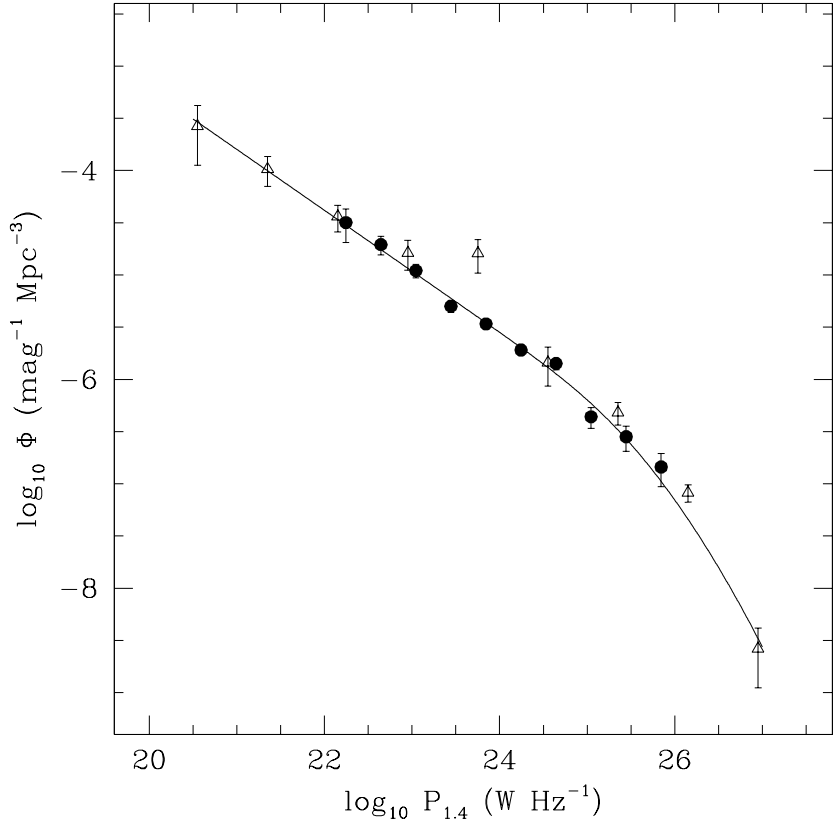}
\caption{Local radio luminosity function for active galaxies at 1.4\,GHz.  Filled circles are points from the 2dFGRS--NVSS data set, while open 
triangles are the Sadler et al.\  (1989) values for 
nearby E/S0 galaxies, converted from 5\,GHz assuming a spectral index of $\alpha=-0.7$.  The solid line shows the best--fitting value of an 
analytic fitting function as described by Saunders et al.\ (1990).   
Luminosities are calculated using $H_0$ = 50 km s$^{-1}$ Mpc$^{-1}$ 
and $\Omega_0$=1. }
\end{figure}

Figure 1.1 shows the local radio luminosity function for active galaxies 
at 1.4\,GHz, as derived by Sadler et al.\ (2002). The values are in 
good agreeement with earlier determinations, but have much lower 
error bars in the luminosity range 10$^{22}$ -- 10$^{25}$ W\,Hz$^{-1}$. 
Remarkably, the space density of radio--emitting AGN is extremely close 
to a simple power--law of the form 
\begin{equation}
\Phi(P_{1.4}) \propto P_{1.4}^{-0.62\pm0.03}
\end{equation}
over almost five decades in luminosity from $10^{20.5}$ to $10^{25}$ 
W\,Hz$^{-1}$, before turning down above $10^{25}$ W Hz$^{-1}$.  

\section{The Local Black Hole Mass Function for Early-type Galaxies} 

Following the precepts of Franceschini et al.\ (1998), the 1.4\,GHz 
radio luminosity function in Figure 1.1 can be converted to a black 
hole number density assuming 
\begin{equation}
\log_{10} M_{\rm BH}\ (M_\odot) = 0.376\ \log_{10} P_{1.4}\ 
({\rm W\,Hz}^{-1}) + 0.173.  
\end{equation}
 
Figure 1.2 shows the results --- the downturn in the radio luminosity function 
above 10$^{25}$ W\,Hz$^{-1}$ implies a corresponding steeper downturn 
in the space density of black holes more massive than about 
$5\times10^9$\,M$_\odot$.  The derived number densities mostly agree 
well with the values derived by Franceschini et al.\ (1998), though the 2dFGRS--NVSS points are systematically higher at the high--mass end.  
This may be the result of luminosity evolution within the relatively 
large volume (to $z\sim0.2$) probed by the 2dFGRS for powerful radio sources.  

Figure 1.3 shows the mass density distribution, 
i.e.\ the number density of black holes of a given mass multiplied by 
that mass. Again, there is a stronger downturn above a few times 10$^9$ 
solar masses.  At the low--mass end, the mass density of black holes 
continues to increase down to the lowest values (a few times $10^7$ M$_\odot$) probed by radio surveys. Integrating over the curve in Figure 1.4 
gives the total mass density of massive black holes 
($M_{\rm BH} > 7.6\times10^7$\,$M_\odot$) in nearby galaxies:  
\begin{equation}
\rho_{\rm BH}\ =\ 1.8^{+0.4}_{-0.6}\times10^5\ M_\odot\ {\rm Mpc}^{-3} 
\end{equation}
  
\noindent
This is clearly a lower limit, since the derived black--hole 
mass density is still increasing at the lowest values of M$_{\rm BH}$ we can measure.  However, our value for $\rho_{\rm BH}$ agrees well with the 
values recently derived by Yu \& Tremaine (2002) and Aller \& Richstone 
(2002) from optical measurements of galaxy velocity dispersions.  

It is also within the range $1.4-2.2\times 10^5$ derived by Chokshi \& Turner 
(1992) from the optical luminosity function of QSOs, implying that local 
radio--emitting AGN are the direct descendants of most or all of the 
high--$z$\ QSOs.  Similar conclusions have been reached by others 
(e.g. Franceschini et al. 1998; Salucci et al.\ 1999; Merritt \& Ferrarese 
2001).  

\begin{figure}
\centering
\includegraphics[width=\columnwidth]{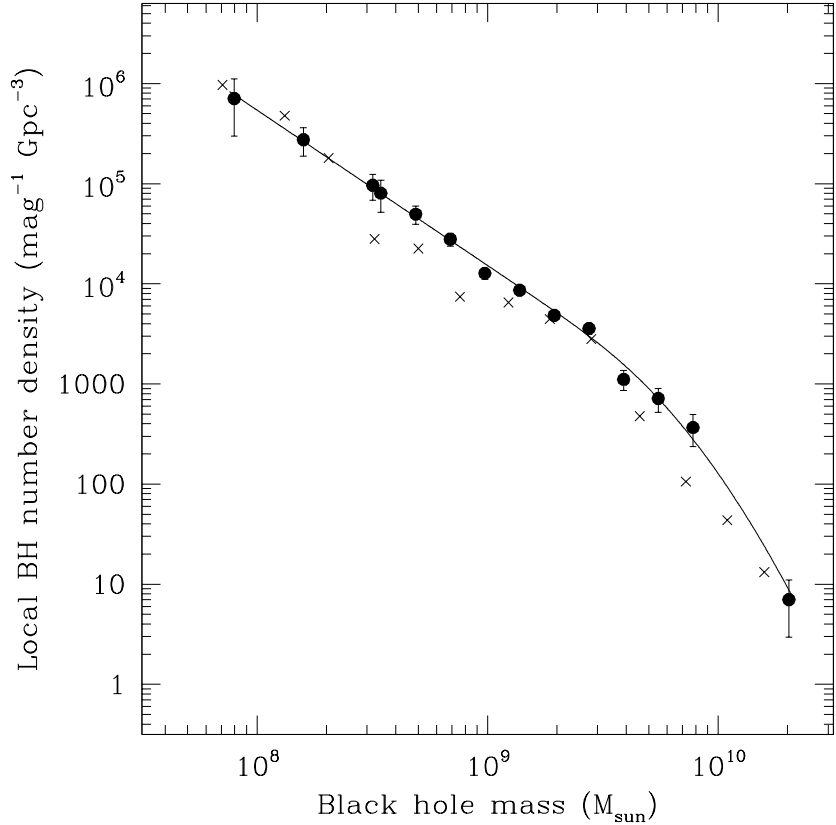}
\caption{Black hole number density, in mag$^{-1}$\,Gpc$^{-3}$ for radio 
galaxies in the local universe.  The filled circles and solid line are 
converted from the data points and analytic fit shown in Figure 1.1.  
Crosses show values from Franceschini et al.\ (1998).  }
\end{figure}

\begin{figure}
\centering
\includegraphics[width=\columnwidth]{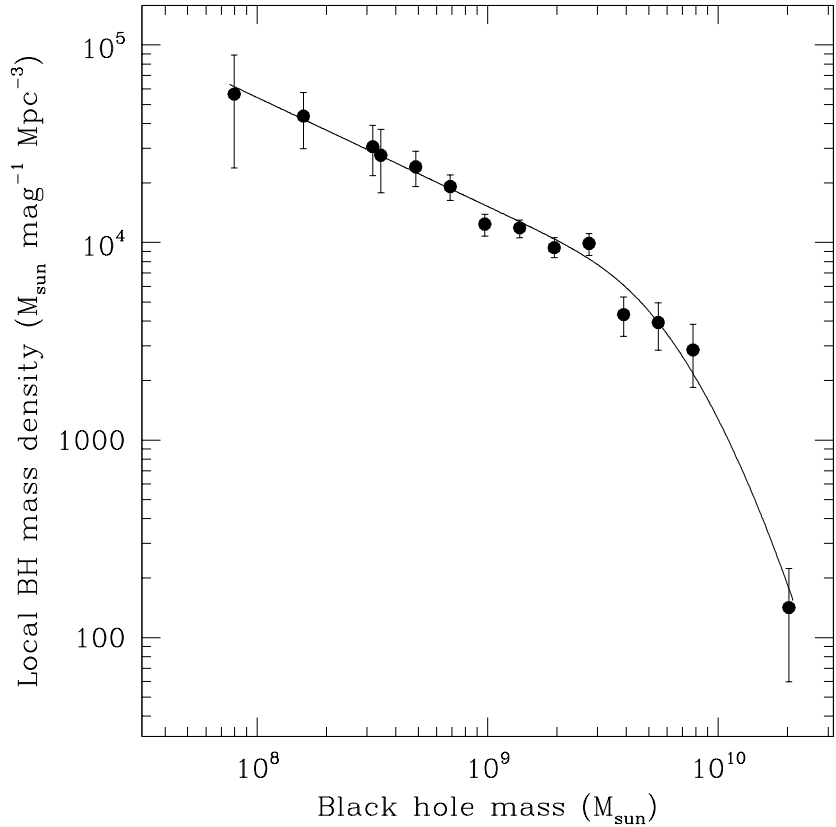}
\caption{Local black--hole mass density (in M$_\odot$\,mag$^{-1}$\,Mpc$^{-3}$).   As in Figure 1.2, the filled circles and solid line 
correpond to the data points and fit from Figure 1.1.  }
\end{figure}

\section{Conclusions} 
Both massive black holes and low--power radio sources are  
common in nearby elliptical galaxies.  If radio luminosity is a good 
estimator of black--hole mass in these galaxies, then a comparison of the 
black--hole mass density in nearby early--type galaxies and distant QSOs 
suggests that local radio-emitting AGN (i.e.\ luminous elliptical galaxies) 
are the direct descendants of most or all of the distant QSOs.  \\

\noindent
{\bf Acknowledgements} 

\noindent
Much of this work was done in collaboration with members of the 2dFGRS 
team, and is described in more detail in our joint paper.

\end{document}